\documentclass[lettersize,journal]{IEEEtran}
\usepackage{amsmath,amsfonts}
\usepackage{algorithmic}
\usepackage{algorithm}
\usepackage{array}
\usepackage[caption=false,font=normalsize,labelfont=sf,textfont=sf]{subfig}
\usepackage{textcomp}
\usepackage{stfloats}
\usepackage{url}
\usepackage{verbatim}
\usepackage{graphicx}
\usepackage{cite}
\usepackage{braket}

\begin{document}

\title{Information-driven Nonlinear Quantum Neuron}

\author{Ufuk Korkmaz and Deniz Türkpençe
\thanks{The authors acknowledge support from the Scientific and Technological Research Council of Turkey (TÜBİTAK-Grant No. 120F353). The authors also wish to extend special thanks to the  Cognitive Systems Lab in the Department of Electrical Engineering providing the atmosphere for motivational and stimulating discussions.(Corresponding author:Ufuk Korkmaz)}
\thanks{Manuscript received January 30, 2023; revised ???? ??, 2023. The authors are with the Department of Electrical Engineering, İstanbul Technical University, İstanbul,
, 34469 TURKEY e-mail: (ufukkorkmaz@itu.edu.tr; dturkpence@itu.edu.tr}}

\markboth{Journal of \LaTeX\ Class Files,~Vol.~??, No.~??, January~2023}%
{Shell \MakeLowercase{\textit{et al.}}: A Sample Article Using IEEEtran.cls for IEEE Journals}



\maketitle

\begin{abstract}
The promising performance increase offered by quantum computing has led to the idea of applying it to neural networks. Studies in this regard can be divided into two main categories: simulating quantum neural networks with the standard quantum circuit model, and implementing them based on hardware. However, the ability to capture the non-linear behavior in neural networks using a computation process that usually involves linear quantum mechanics principles remains a major challenge in both categories. In this study, a hardware-efficient quantum neural network operating as an open quantum system is proposed, which presents non-linear behaviour. The model's compatibility with learning processes is tested through the obtained analytical results. In other words, we show that this dissipative model based on repeated interactions, which allows for easy parametrization of input quantum information, exhibits differentiable, non-linear activation functions.
\end{abstract}


\section{Introduction}

\IEEEPARstart{A}{rtificial} neural networks (ANNs) play a crucial role in machine learning.
Complex tasks have been achieved with artificial neural network models in the past two decades, especially with the increase in the capacities of computer processors~\cite{arulkumaran_deep_2017,mahmud_applications_2018,fadlullah_2017,silver_mastering_2016}. However, the increasing volume of data to be processed and the reaching of the end of Moore's law pose significant challenges to the rate of performance increase of artificial neural networks~\cite{markov_limits_2014}.

Quantum computing is a computational paradigm harnessing resources that do not have classical counterparts~\cite{nielsen_quantum_2011}. With only a few algorithms presented by this paradigm, it has been demonstrated that solving some problems believed to be impossible with classical computation methods is possible~\cite{shor_polynomial-time_1997, van_dam_quantum_2006, hallgren_polynomial-time_2007}. Due to the reasons mentioned above, proposals for machine learning and artificial neural network models based on the existing advantages of quantum computing have begun to emerge in parallel with the growing interest in quantum computing~\cite{harrow_quantum_2009, lloyd_quantum_2014,biamonte_quantum_2017,  cong_quantum_2019}. Despite various proposals regarding quantum artificial neural networks~\cite{schuld_quest_2014, massoli_leap_2022}, the lack of agreement on a widely accepted model has made the topic an open field of research. Particularly at the very fundamental level, the problem of simulating the non-linearity of a quantum neuron with the widely linear properties of quantum computing is a major issue that needs to be resolved.

The implementation proposals for artificial neural networks~\cite{zhao_building_2019,monteiro_quantum_2021} and non-linear quantum neurons~\cite{yan_nonlinear_2020,de_paula_neto_implementing_2020} on standard quantum circuit models often require high resource costs. In addition, if multi-controlled gates for the neuron gate are required in the algorithm specifically prepared for the problem to be solved, the need for time-dependent optimization to construct these gates has the potential to limit the efficiency of the algorithm. Moreover, the current quantum computers being constructed are largely influenced by quantum noise, which severely limits the efficiency of quantum algorithms requiring high resources. This era, referred to as the Noisy Intermediate Scale Quantum computing~\cite{preskill_quantum_2018} (NISQ), is characterized by this phenomenon and efforts are being made to address it~\cite{bharti_noisy_2022}. The dissipation-driven quantum computing model was introduced as an equivalent to the standard circuit quantum computing model~\cite{verstraete_quantum_2009}. Thus, the dissipation-driven model offers robustness against quantum noise and is also a standalone analog quantum computing model.

In this study, we introduce a dissipation-assisted quantum neuron model with non-linear response that operates as a binary classifier quantum perceptron, where the binary decision is read out by a probe quantum system (PQS) with spin angular momentum $J\geq 1/2$. In the model, the PQS is in contact with multiple, distinct quantum environments in pure quantum states with arbitrary coupling rates as input quantum data. In the scenario, PQS undergoes a dissipative equilibration process and reaches a steady state where the binary decision is encoded. Steady state magnetization is the merit quantifier in our model. The dissipation process is characterized by a collision model based on repeated interactions~\cite{scarani_thermalizing_2002,ziman_diluting_2002,cattaneo_collision_2021,ciccarello_quantum_2022}, which effectively characterizes open quantum systems and enables easy parametrization of input quantum information contained by the reservoirs. The environments bearing quantum information is referred to as information reservoirs~\cite{deffner_information_2013,deffner_information-driven_2013}. 

We show that the quantum neuron responds as a hyperbolic tangent-like activation which can be controlled by the value of $J$. We derive a master equation and analytically obtain the steady-state solutions. We also derive a cost function based on the obtained results and examine the performance on gradient descent-based learning tasks. We observe that the proposed neuron is suitable for learning schemes as the dissipation based process results a continuous dynamics with proper gradient minimization. Finally, the steady response of the neuron, driven by non-equilibrium environments, contains non-vanishing quantum coherence. This opens up the possibility for future work to harness non-classical quantum resources in a steady state. 

\section{Preliminaries}\label{msd}

As discussed above, proposals related to quantum models of artificial neural networks are still at the level of basic discussions, so we aim to start with definitions from the most basic classical model and thus establish a connection with the quantum model we propose. 

\subsection{The classic model}
The learning theory of artificial neural networks is based on mathematical models that mimic the functioning of the human brain, as first proposed by McCulloch, Pitts, and Rosenblatt~\cite{mcculloch_logical_1943,rosenblatt_perceptron_1958}. 
The perceptron is a foundational component of artificial neural networks carrying out binary classification tasks. In essence, it outputs a binary label of either 0 or 1 based on the input, which is represented as $\varphi_{in}=\bf{x}^T \bf{w}$, where $\bf{x}=[x_1,\ldots x_N]^T$ is the set of input features and $\bf{w}=[w_1,\ldots w_N]^T$ the corresponding weight vectors. This binary output is generated through a non-linear function $f(.)$, where $z=f(\varphi_{in})$, and the decision rule is set as $z=0$ if $z=f(\varphi_{in})\geq 0$ and $z=1$ otherwise. The specific binary labels can vary as needed. It's important to note that a perceptron with an identity activation function can still produce linear classification, though using non-linear activation functions are beneficial for multi-layer ANN applications.

Supervised learning involves creating a mapping from input data to binary labels
\begin{equation}\label{Eq:Mapping}
\mathcal{X,Y}\rightarrow \{0,1\}.
\end{equation}
The input data, represented as $\mathcal{X}$, and the desired output, represented as $\mathcal{Y}$, are part of a training set $\mathcal{S}=(\mathcal{X},\mathcal{Y})$. The cost function, $C$, measures the difference between the actual output vector, $\bf{A}$, and the desired output vector, $\bf{Y}$. The cost function expression follows the least squares method, as 
\begin{equation}\label{Eq:Cost1}
C=\frac{1}{2}(\bf{Y}-\bf{A})^2.
\end{equation} 
The weight instances are updated iteratively through back-propagation 
\begin{equation}\label{Eq:Weight}
\bf{w}_{k+1}:=\bf{w}_{k}+\delta\bf{w}_{k}
\end{equation}
to minimize the cost. A gradient-descent based method is used for the training process, and the change in the parameters is determined by the partial derivative, as 
\begin{equation}\label{Eq:delta_w}
\delta {w}_k=-\eta\frac{\partial C}{\partial {w}_k}.
\end{equation}
The learning rate, represented by $\eta$, influences the speed of the learning task and is non-negative. 
The partial derivative expresses the change in the parameter that leads to the largest descent.

\subsection{The quantum dynamics}

This section addresses the dynamics of open systems and gives preliminary definitions. Our model operates based on a dissipative protocol as previously noted. The input data, presented in a classical manner, is represented as the weighted sum of input features 
\begin{equation}\label{Eq:Weight}
\varphi_{in}=\bf{x}^T \bf{w}=\sum_i w_i x_i. 
\end{equation}
We describe the dissipative quantum equivalent of the above expression as 
\begin{align}\label{Eq:CPTP addition}
\Lambda_t[\varrho_0]=\sum_i P_i\Phi_t^{(i)}{[\varrho_0] }
\end{align}
where $\Phi_t^{(i)}$ is a completely positive trace preserving (CPTP) quantum dynamical map that acts on the probe system characterized by a density matrix $\varrho_0$ and $P_i$ represents the probability of the map interacting with the $i$th information reservoir. The subscript $t$ in Eq.~(\ref{Eq:CPTP addition}) represents the time-dependence of the maps, which are generated by a physical process 
\begin{align}\label{Eq:Unitary}
\Phi^{(i)}_t[\varrho_0]=\text{Tr}_{\mathcal{R}_i}\{U_t(\varrho_0\otimes\varrho_{\mathcal{R}_i})U_t^{\dagger}\}
\end{align}
that has a unitary propagator $U_t$ that acts on both the PQS and the reservoir. The $i$th reservoir quantum state is represented by $\rho_{\mathcal{R}i}$, and the partial trace over the $i$th reservoir is represented by $\text{Tr}_{\mathcal{R}_i}$.

The quantum reservoirs provide initial quantum data, each composed of non-correlated, non-interacting two-level quantum systems (subunits) defined by the tensor product 
\begin{align}\label{Eq:Inf Res}
\rho_{\mathcal{R}_i}=\bigotimes_{k=1}^n\rho_{k}(\theta_i,\phi_i).
\end{align}
These subunits are initially prepared in pure quantum states with identical Bloch parameters $\rho_{k}(\theta_i,\phi_i)$, which enables the dissipative equivalence of the model with parametrized quantum circuits.

\section{Collision model and the quantum neuron}

The model described relies on a standard quantum collisional model for its dynamics. The probe system experiences a dissipative process due to multiple independent reservoirs with arbitrary couplings. The steady state of the probe is read through spin observable $S_z$, providing a binary classification output. The dynamics are given by 
\begin{align}
\Phi^{(i)}_{n\tau}=&\text{Tr}_n \big[ \mathcal{U}_{i_n}\ldots\text{Tr}_1[\mathcal{U}_{i_1}\left(\varrho_0\otimes\rho_{\mathcal{R}_{i_1}}\right)\mathcal{U}_{i_1}^{\dagger}]\otimes\ldots \nonumber \\ 
&\ldots\otimes\rho_{\mathcal{R}_{i_n}}\mathcal{U}_{i_n}^{\dagger} \big].
\end{align}   
where $n\tau$ is the elapsed time of the map after $n$ collisions and $\mathcal{U}_{i_k}=\text{exp}[-\text{i}\mathcal{H}^k_{i}\tau]$ is the unitary propagator applying the $k$th collision to the $i$th reservoir. The system and the $i$th reservoir are initially prepared in a tensor product state $ \varrho (0)= \varrho_0 (0)\otimes\rho_{\mathcal{R}_i} $. Only the probe system experiences time dependence, and the reservoir states are reset after each collision.

The dynamics of the system and reservoir are governed by the Hamiltonian $\mathcal{H}= \mathcal{H}_0+\mathcal{H}_{int}$, where 
\begin{align}
\mathcal{H}_0=\hbar\omega_0 S_0^z+\frac{\hbar\omega_i}{2}\sum_{k=1}^n\sigma_{k_i}^z
\end{align}\label{eq:hfree}
is the free part and
\begin{equation}
\mathcal{H}_{int}=\hbar\sum_{k=1}^n g_i(S_{0}^+\sigma_{k_i}^- +\text{H.c.}),
\label{eq:hint}
\end{equation}
is the interaction part. Respectively, the PQS raising operator, Pauli-$z$ operator, Pauli raising and lowering operators are represented as $S_0^+$, $\sigma^z$, $\sigma^+$, and $\sigma^-$, with $\hbar$ Planck's constant divided by $2\pi$ set to 1. The coupling coefficients $g_i$ are in the weak coupling regime ($g_i\ll \omega_0$), ensuring minimal cross-talk between reservoirs. The coefficients are proportional to probabilities ($g_i\propto P_i$) in Eq.~(\ref{Eq:CPTP addition}), acting as the quantum equivalent of weights in classical models. 

In this section, we present numerical and analytical results. As mentioned above, the output results are read from a quantum system, PQS, that behaves in a spin state $J\geq 1/2$. In quantum mechanics, systems that behave as spin with $J\geq 1/2$ have $2J+1$ orientations relative to the quantization axis. This result suggests that high spin states can be adapted to a multi-class classification problem with observables in our methods. However, this idea requires further study and analysis, so it falls out of the scope of this study where we focus on binary classification.

Regardless of the total $J$ value of the PQS system, it always gives a binary classification result associated with the relevant $S_z$ observables, which are always non-linear with input quantum data weighted by the coefficients $g_i$. We use the methods from our previous study~\cite{korkmaz_quantum_2023}, where the derivation of the master equation and the analytical results were limited to $J=1/2$ and only gave linear activation results. In this study, we present how PQS behaves like a non-linear activation function for $J\geq 1/2$ cases and its effects on minimizing the cost function. In addition, we discuss how PQS can be physically applied for $J\geq 1/2$ in a separate section at the end of the study. 

For any spin value $J$, the corresponding generalized density matrix expression can be defined through the basis of polarization operators. The density operator expression for PQS, therefore, reads
\begin{align}\label{Eq:SpinRho}
\varrho_0& =\frac{1}{d}\bf{1}+\overrightarrow{r}\cdot\overrightarrow{T}\nonumber\\
&=\frac{1}{d}\bf{1}+\sum_{l=1}^{2J}\sum_{m=-l}^l r_{lm}T_{lm}
\end{align}
where $\bf{1}$ is the d-dimensional identity operator, $\overrightarrow{r}$ is the generalized Bloch vector whose elements are $r_{lm}=\text{Tr}T_{lm}^{\dagger}\varrho_0$ and $\overrightarrow{T}$
is the relevant spin polarization vector. Alternative generalized density matrix representations are also possible through the generalized Gell–Mann matrix basis or the Weyl operator basis~\cite{bertlmann_bloch_2008}. However, we will present our analytical results below through polarization operator basis representation.

\subsection{Analytical results}
The unitary propagator is expressed as $\mathcal{U}(\tau)=\text{exp}[-\text{i}\mathcal{H}_{\text{int}}\tau]$ in the interaction picture (as detailed in Appendix A) with respect to $\mathcal{H}_0$. This is used to derive the master equation,
\begin{align}\label{Eq:master eq}
\dot{\varrho}_0=&-\text{i}[\mathcal{H}_{\text{eff}},\varrho]+\sum_{i=1}^N g_i^2\left(\xi^{+}\mathcal{L}[S_0^{+}]+\xi^{-}\mathcal{L}[S_0^{-}]\right)
\nonumber\\
&+\sum_{i<j}^{N'} g_i g_j \left(\xi^{+}_s \mathcal{L}_s[S_0^{-}]+\xi^{-}_s \mathcal{L}_s[S_0^{+}]\right)
\end{align}
where $\mathcal{H}_{\text{eff}}=r\tau\sum_i^N g_i\left(\langle\sigma_i^{-}\rangle S_0^{+}+\langle\sigma_i^{+}\rangle S_0^{-}\right)$, representing a coherent drive on the probe qubit as the effective Hamiltonian. The averages are calculated for identical reservoir units, i.e. $\langle \mathcal{O}i\rangle=\text{Tr}[\mathcal{O}\varrho_{\mathcal{R}_i}]$. The Lindblad superoperator is defined as $\mathcal{L}[o]\equiv 2o\varrho o^{\dagger}-o^{\dagger}o\varrho-\varrho o^{\dagger}o$, and $\mathcal{L}_s[o]\equiv 2o\varrho o -o^2\varrho-\varrho o^2$ refers to the effect of squeezing by the reservoir. The coefficients of the Lindbladians contain information related to different entries of the density matrices of the reservoir units. For example, the standard Lindbladian coefficients $\xi^{\pm} = r\tau^2\langle \sigma_i^{\pm}\sigma_i^{\mp}\rangle/2$ include the diagonal entries of the $i^{th}$ reservoir unit, while $\xi_s^{\pm} = 2r\tau^2\langle \sigma_i^{\pm}\rangle\langle\sigma_j^{\pm}\rangle$ includes the off-diagonal entries of the pairwise distinct reservoir units, with a total of $N' = N(N-1)/2$ terms in the summation. Here, $r$ is a positive number used in forming the master equation and represents the success of any collision (see Appendix B).

Based on the model we presented, we obtained the steady state solution of the master equation, since the quantum neuron binary classification decision is given in the steady state. The steady state solution of Eq.~(\ref{Eq:master eq}) can be obtained straightforwardly for any $J$ by using the related spin operators $S_0^{\mp}$. However, as the analytical results become too cumbersome to generalize the result, we deal with the case for $J=1$ analytically and evaluate the higher spin cases numerically. Therefore the steady state of PQS reads 
\begin{equation}\label{Eq:Steadyy} 
\varrho_0^{\text{ss}}=\begin{bmatrix} P_{11} & \gamma^-_1(P_{11}-P_{22}) & 2\gamma^-_3P_{22} \\
\gamma^+_1(P_{22}-P_{11}) &  P_{22} & \gamma^-_1(P_{22}-P_{33}) \\
2\gamma^+_3P_{22}  & \gamma^+_1(P_{33}-P_{22}) & P_{33} \end{bmatrix}
\end{equation}
where $P_{nn}$ are the steady state populations (see Appendix B for detailed expressions) and $\gamma^{\pm}_1=\text{i}r\tau\sum_{i=1}^N g_i\langle\sigma_{i}^{\pm}\rangle$,$\gamma^{\pm}_2=r\tau^{2}\sum_{i=1}^N g^2_i\langle\sigma_{i}^{\pm}\sigma_{i}^{\mp}\rangle/2$, $\gamma^{\pm}_3=2r\tau^2\sum_{i<j}^{N'}g_i g_j \langle\sigma_{i}^{\pm}\rangle\langle\sigma_{j}^{\pm}\rangle$. Please note that the $\xi^{\pm}$ coefficients in Eq.~(\ref{Eq:master eq}) are directly related to the elements of the density matrix of the reservoir units, and the $\gamma^{\pm}$ coefficients in Eq.~(\ref{Eq:Steadyy}) are the weighted summation of these elements. These results analytically prove that PQS in its steady state gives the weighted sum of the input quantum data. We will evaluate the following numerical results for the activation behaviour.

\begin{figure*}[t]
\centering
\subfloat[]{\includegraphics[width=3.2in]{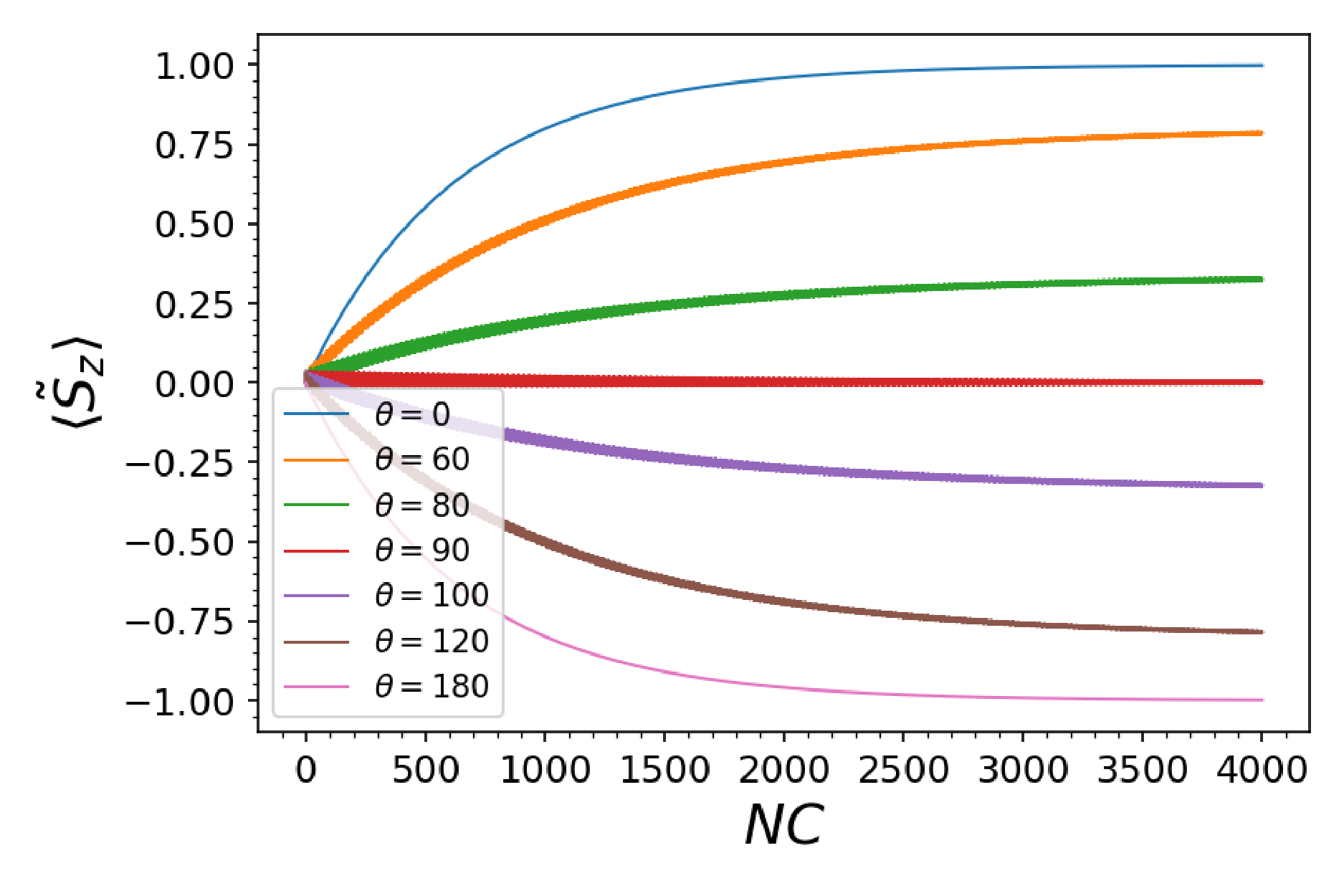}%
\label{fig:sz_nd_1}}
\hfil
\subfloat[]{\includegraphics[width=3.2in]{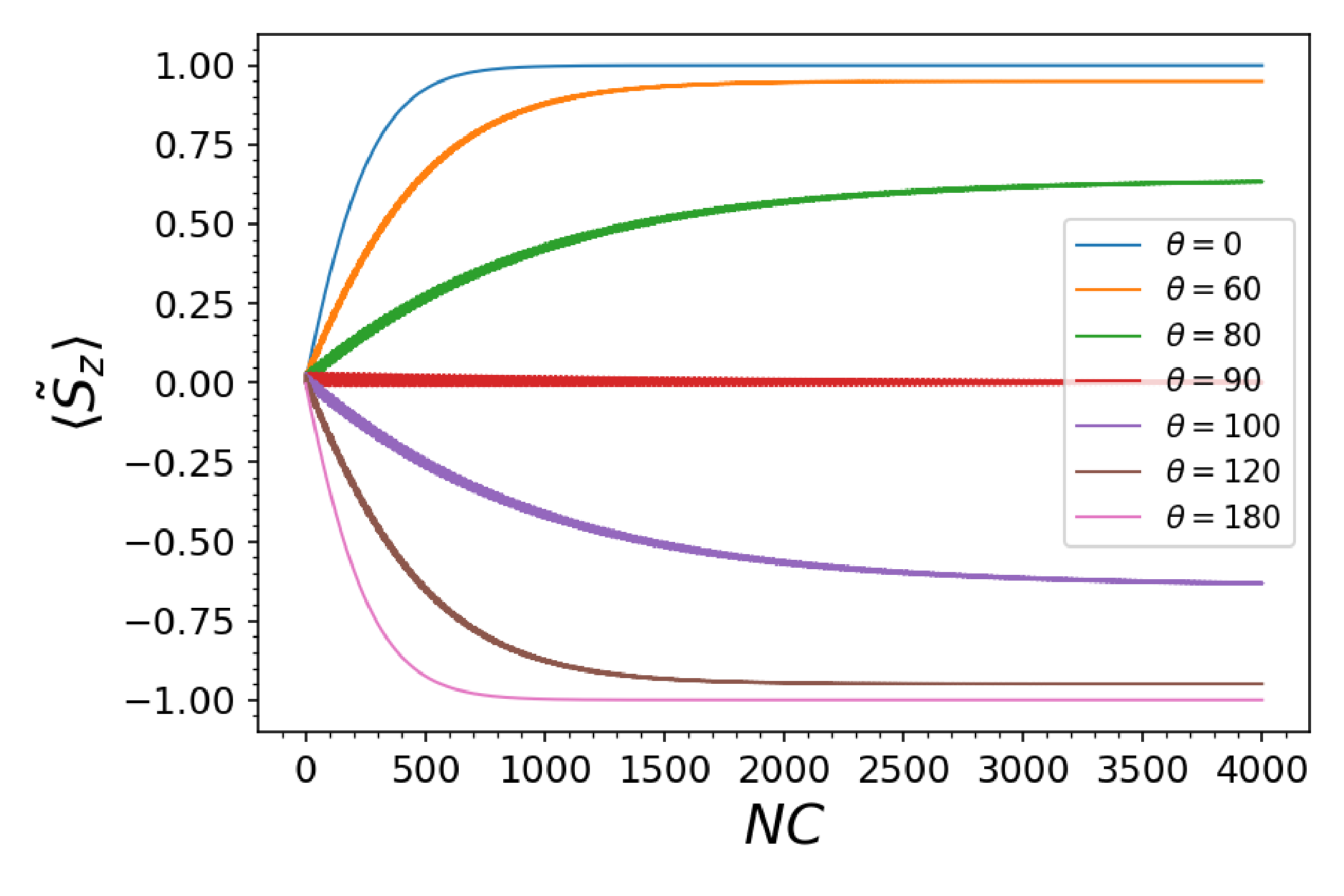}%
\label{fig:sz_nd_5}}
\caption{Equilibration dynamics of the probe qubit magnetization against the collision number depending on the reservoir qubit amplitude parameter $\theta$. 
The probe qubit initialized in the spin-coherent state $\ket{\bf{n}}$. The reservoir qubit initialized as $\rho_{\mathcal{R}}=\ket{\Psi(\theta,\phi)}\bra{\Psi(\theta,\phi)}$ for $\theta=0^{o},60^{o},80^{o},90^{o},100^{o},120^{o},180^{o}$ (from top to bottom) and $\phi=0$. The target qubit-reservoir interaction time $\tau=3$ and the coupling coefficient $g=0.02$ are dimensionless and scaled by $\omega_r=10^9$Hz. (a) The spin angular momentum value is $J=1/2$. (b) The spin angular momentum value is $J=5/2$.}
\label{fig_sim}
\end{figure*}

\begin{figure*}[!t]
\centering
\subfloat[]{\includegraphics[width=3.5in]{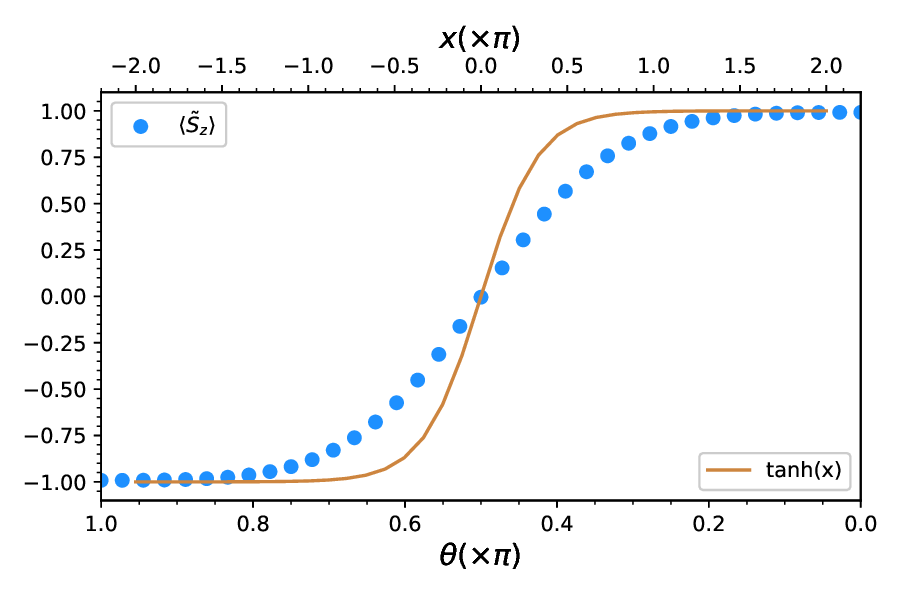}%
\label{fig:Activation_Nd_1_tan}}
\hfil
\subfloat[]{\includegraphics[width=3.5in]{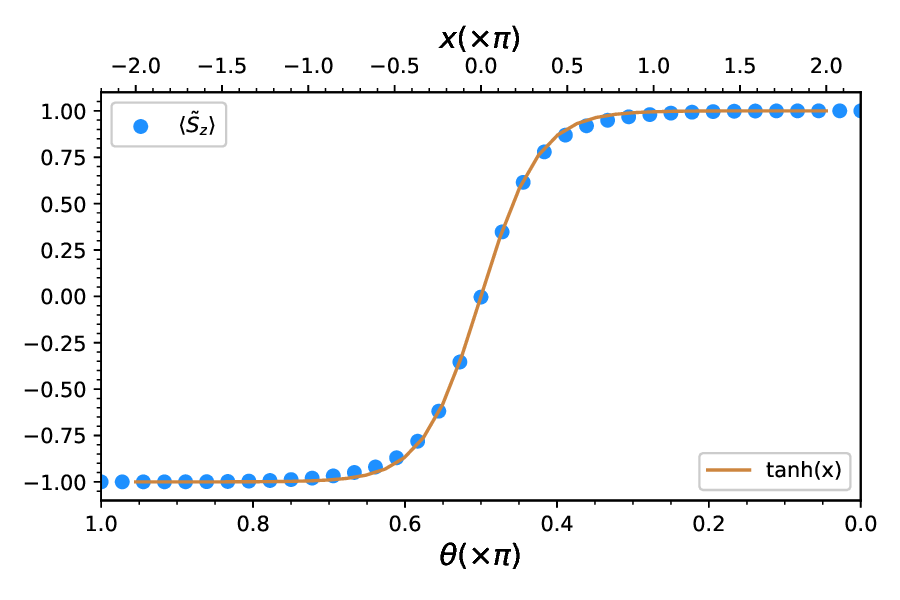}%
\label{fig:Activation_Nd_5_tan}}
\hfil
\subfloat[]{\includegraphics[width=3.5in]{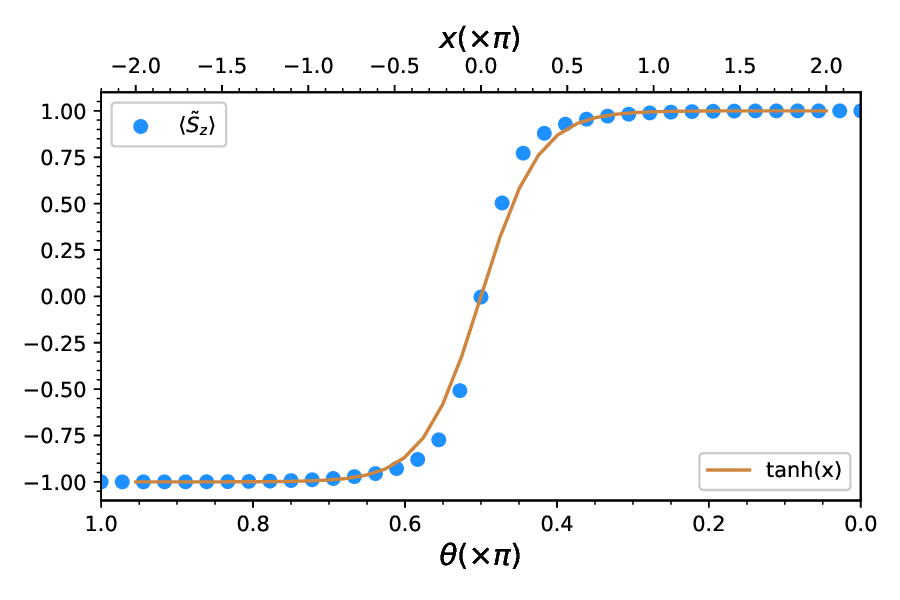}%
\label{fig:Activation_Nd_9_tan}}
\caption{The activation function is in the form of equilibrium dynamics of probe qubit magnetization versus reservoir qubit amplitude parameter $\theta$. 
The probe qubit initialized in the spin-coherent state $\ket{\bf{n}}$. The reservoir qubit initialized as $\rho_{\mathcal{R}}=\ket{\Psi(\theta,\phi)}\bra{\Psi(\theta,\phi)}$ for $\theta=0^{o},60^{o},80^{o},90^{o},100^{o},120^{o},180^{o}$ and $\phi=0$. The target qubit-reservoir interaction time $\tau=3$ and the coupling coefficient $g=0.02$ are dimensionless and scaled by $\omega_r$. (a) The spin angular momentum value is $J=1/2$. (b) The spin angular momentum value is $J=5/2$. In this case, the magnetization of the probe qubit is hyperbolic tangent ($\tanh(x)$). (c) The spin angular momentum value is $J=9/2$.}
\label{fig_sim}
\end{figure*}

\subsection{Numerical results}

The numerical analysis begins by examining the simplest scenario where the PQS is connected to a single (only $g_1\neq 0$) information reservoir in the quantum state $\varrho_1=\ket{\Psi(\theta_1,\phi_1)}\bra{\Psi(\theta_1,\phi_1)}$. The progression of events is monitored by the number of collisions with quantum data parameters represented by $\theta$. Throughout the analysis, $\phi$ is set to 0, without loss of the generalization. The initial state of the PQS is a spin coherent state 
\begin{multline}\label{Eq:coherent}
\ket{\bf{n}}=\sum_{m-j}^j \sqrt{\binom{2j}{j-m}} \left(\cos(\frac{\theta}{2} )\right)^{j+m}\\
\times \left(\sin(\frac{\theta}{2})e^{\text{i}\phi}\right)^{j+m}\ket{j,m}
\end{multline}
with null magnetization where $\ket{j,m}$ are the standard angular momentum basis states, $\bf{n}=(\sin\theta\cos\phi,\sin\theta\sin\phi,\cos\theta)^T$ with $\theta\in[0,\pi]$ and $\phi\in[0,2\pi[$. Here, $m$ is the spin projection number with $m=-J,\cdots J$. The simulations are carried out by using QUTIP package in Python \cite{johansson_qutip_2013} and the parameters are based on the superconducting circuits~\cite{deng_robustness_2017,krantz_quantum_2019,blais_circuit_2021}, which serve as a reliable platform for quantum information processing. 

In this architecture, transmon qubits are connected through a resonator bus~\cite{majer_coupling_2007}, allowing interaction through virtual photon exchange. The coupling strength between qubits can be adjusted by their dispersive connection to the transmission line resonator. A typical superconducting circuit with weakly connected transmon qubits has a resonator frequency of $\omega_r$ ranging from 1-10 GHz with a qubit-resonator coupling of $g_r$ from 1-500 MHz and an effective qubit-qubit coupling of $g$ from 1-100 MHz, with a qubit energy relaxation time of $T_1$ ranging from 40-150 $\mu s$~\cite{deng_robustness_2017,majer_coupling_2007}.

Fig.~\ref{fig:sz_nd_1}  illustrates the temporal evolution of PQS (for $J=1/2$) as a function of the number of collisions ($NC$) in the presence of varying $\theta$ parameters of a single information reservoir. The monitored quantity is the normalized magnetization, defined as $\langle\tilde{S}_z\rangle=\text{Tr}[\varrho(n\tau)S_z]/J$. It is noted that PQS reaches a steady state after $n\sim 2\times 10^3$ collisions. With an interaction time of $\tau=3$ ns, the time required to reach the steady state is $n\tau \approx 6 \mu$s, which is much smaller than the $T_1$ value.
It has been reported that the time required to reach the steady state decreased when PQS interacted with environments containing more than a single qubit~\cite{turkpence_tailoring_2019}. By keeping all parameters constant except for the $J$ parameter and setting $J=5/2$, as seen in Fig.~\ref{fig:sz_nd_5}, all the $\langle\tilde{S}_z\rangle$ values corresponding to the $\theta$ parameters of the information reservoir exhibit a non-linear behaviour and approach the steady state values of $\langle\tilde{S}_z\rangle=1$ and -1. In principle, single-input quantum data can still be classified. While we continue to analyse the model with a single information reservoir, a study of the model's behaviour in the presence of multiple information environments will be performed. 

To gain a comprehensive understanding of the behaviour of the proposed model, alongside a temporal evaluation, we focus on examining the steady state values with respect to $\theta$, as shown in Fig. 2. The steady state magnetization values of PQS marked as dots in the figure correspond to different values of the $J$ parameter, resembling the $tanh(x)$ activation function. For a better comparison, we also added the $tanh(x)$ function represented by a continuous line in the figure. As known, $\theta$ is a geometric parameter belonging to the Bloch sphere that characterizes the input quantum data and can take values between 0 and $\pi$. The $tanh(x)$ function takes any real $x$ value and outputs values of 1 and -1. Therefore, these two functions cannot be directly compared, but a relative comparison can be made with properly established scales. To this end, Fig.~2 is established for two different scales such as $x\in [-2\pi,2\pi]$ and $\theta\in[\pi,0]$. The starting value of $\theta=\pi$ is a conscious choice for convenient comparison of $\langle\tilde{S}_z\rangle$ with $tanh(x)$. 

In Fig.~\ref{fig:Activation_Nd_1_tan}, where PQS is expressed as $J=1/2$, we observe the linear behaviour of the magnetization over a wide range of values. Although this behaviour is sufficient for the model to perform linear classification tasks on its own, it may be insufficient to model complex, non-linear relationships in multi-layered networks established with this parameter. On the other hand, as seen in Fig.~\ref{fig:Activation_Nd_5_tan}, the steady state response of the model for the PQS value of $J=5/2$ is seen to coincide with the hyperbolic tangent function, $tanh(x)$, at the given scales. As seen in Fig.~\ref{fig:Activation_Nd_9_tan}, for higher $J$ values, the system response is seen to converge more strongly towards the values of 1 and -1.

Hyperbolic tangent is often used as a differentiable activation function in artificial neural networks, where the gradient can easily be calculated~\cite{Complexity, apicella_survey_2021}. However, $tanh(x)$ saturates at large positive and negative values, making it challenging to train deep neural networks. This issue is known as the vanishing gradient problem, where the gradients become very small, making it challenging for the optimization algorithm to update the weights. The behaviour of the proposed model being similar to the aforementioned function suggests that it may have similar advantages and disadvantages. Nevertheless, note that the proposed model has the capacity to control its steady behaviour through adjusting the $J$ parameter. Moreover, the analytical results indicate that regardless of the number of quantum input data, the additivity of quantum dynamical maps can be ensured through considering the convex structure of the density matrix formalism. As a result, the model will consistently act as if it is receiving input information from a single reservoir with an effective $\theta$ value always between 0 and 2$\pi$.
\begin{figure}[!t]
\centering
\includegraphics[width=3.5in]{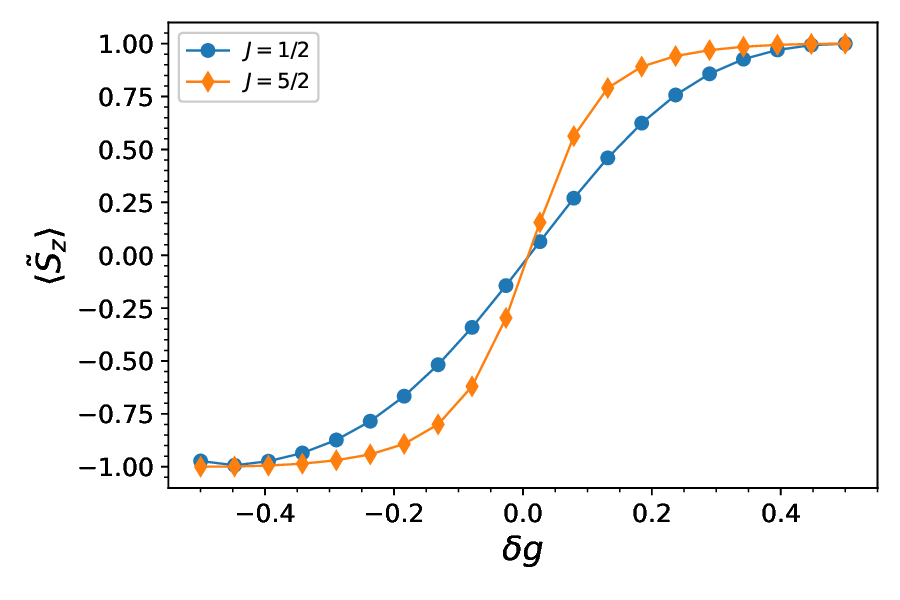}
\caption{The steady state magnetization of the target qubit depending on the variation of the couplings to the reservoirs. The spin angular quantum number is $J=1/2$ for the blue line, and $J=5/2$ for the orange line. Variations of the coupling rates are $g_1=g/2 + \Delta g$, $g_2=g/2 - \Delta g$. Here, $\Delta g$ represents a fraction of $g$. The probe qubit prepared initially in the spin-coherent state $\ket{\bf{n}}$ and interacted collisionally with the reservoir units $\ket{\Psi(\theta_i,\phi_i)}$ with $\theta_1=0$, $\phi_1=0$ and $\theta_2=\pi$, $\phi_2=0$. The target qubit-reservoir interaction time $\tau=3$ and the coupling coefficient $g=0.02$ are dimensionless and scaled by $\omega_r$.}
\label{fig:curve_j}
\end{figure}

After observing the non-linear behaviour of a single input quantum reservoir, we examine the simplest case of multiple reservoirs. Fig.~\ref{fig:curve_j} shows the normalized steady magnetization of PQS with varying $\delta g$. The PQS is connected to two distinct quantum reservoirs, $\ket{\theta_1=0,\phi_1=0}\equiv\ket{\uparrow}$ and $\ket{\theta_2=\pi,\phi_2=0}\equiv\ket{\downarrow}$. The coupling strengths of PQS to each reservoir are $g_1=g/2+\delta g$ and $g_2=g/2-\delta g$ with $g=0.01$ and $-0.5\leq\delta g\leq 0.5$. In the limit case where $\delta g=-0.5g$, $g_1=0$ and $g_2=g$, meaning PQS is only connected to the $\ket{\downarrow}$ reservoir. In this case, as seen in Fig.~\ref{fig:curve_j}, the magnetization is $\langle\tilde{S}_z\rangle=-1$ as expected. For intermediate values of $\delta g$ within the given range, the monitored magnetization varies as $-1<\langle\tilde{S}_z\rangle<1$, and it reaches its other limit value $\langle\tilde{S}_z\rangle=1$ for $\delta g=0.5$. From now on, we will neglect the tilde symbol we use to indicate magnetization for the sake of simplicity.

Thus, we have also shown how the PQS output behaves with multiple quantum information inputs weighted by $g_i$. We also note in Fig.~\ref{fig:curve_j} that the case we studied exhibits different non-linear behaviours for $J=1/2$ and $J=5/2$ situations as a confirmation of Fig. 2. If we compare Fig.~2 and Fig.~3 further, in Fig.~2, where PQS is only connected to a single reservoir, the altered parameter is necessarily $\theta$, whereas in Fig.~3, where PQS is connected to two reservoirs and $\theta$ values are fixed, the $g_i$ values corresponding to the weights against the classical cases change. The behaviour of our output quantifier, magnetization, gives similar results for both cases as the parameters change, indicating that our model is suitable for the training and learning tasks we will discuss below.

\begin{figure*}[!t]
\centering
\subfloat[]{\includegraphics[width=3.2in]{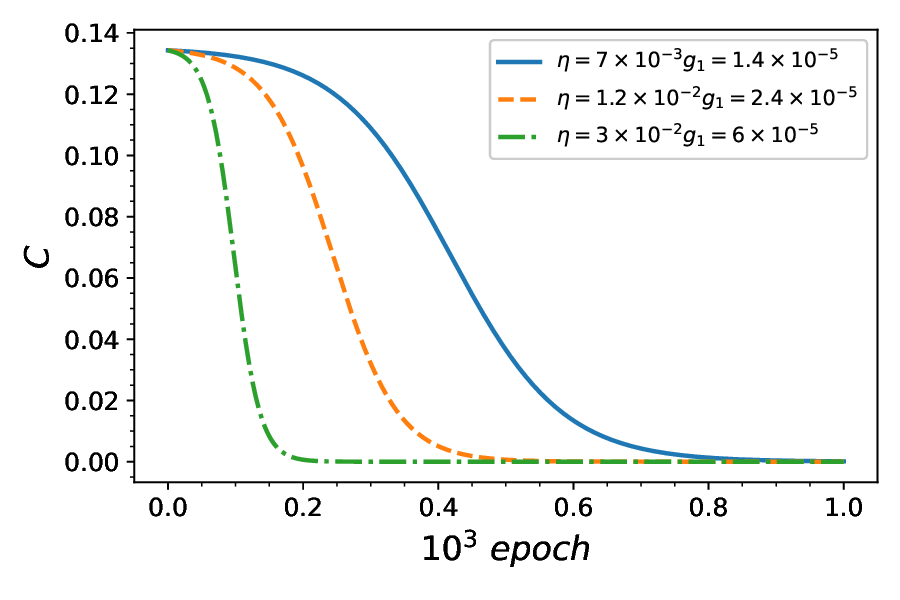}%
\label{fig:Epoch_Nd_1_tan}}
\hfil
\subfloat[]{\includegraphics[width=3.2in]{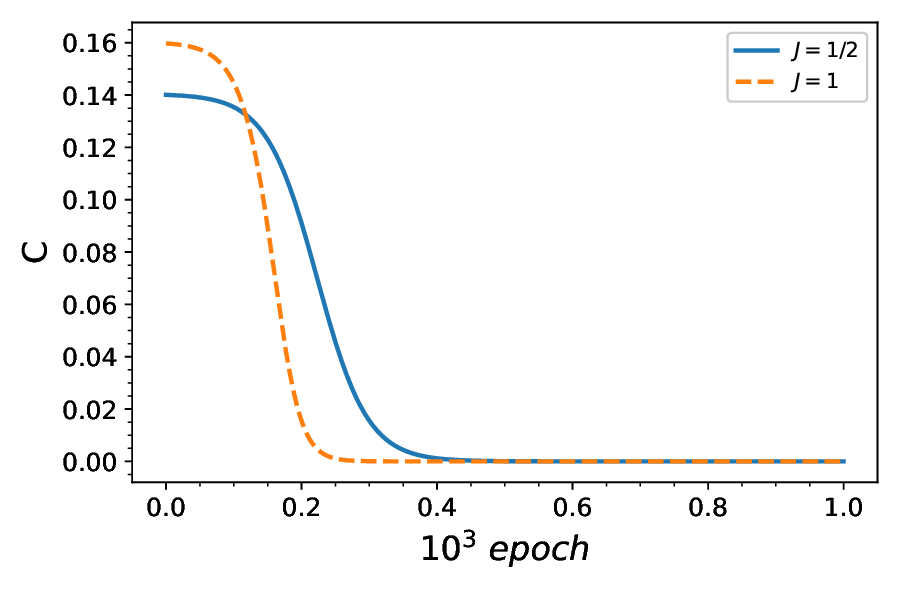}%
\label{fig:Epoch_Nd_1_2}}
\caption{Cost function minimization depending on learning rates against the variation of $g$.(a) The initial magnetization values of the reservoirs are $\langle \sigma_z^1\rangle=0.94$, $\langle \sigma_z^2\rangle=-0.10$, $g_1=0.002$, $g_2=0.05$ and the desired PQS magnetization $\langle S_z^0\rangle_{des}^{ss}=0.42$, respectively. The spin angular momentum value of the PQS is $J=1/2$. (b) The spin angular quantum numbers are $J=1/2$ for the blue solid line, $J=1$ for the orange dotted line. Both of the the initial magnetization values of the reservoirs are  $\langle \sigma_z^1\rangle=0.95$, $\langle \sigma_z^2\rangle=-0.11$, $g_1=0.001$, $g_2=0.04$ and $\langle S_z^0\rangle_{des}^{ss}=0.42$, respectively. The learning rate is $\eta=1.2\times10^{-2} g_1$}
\label{fig_epoch}
\end{figure*}

\section{Learning of the model}

In what follows, we examine the gradient-descent based learning scheme of the introduced model. 
First, we develop the cost function as~\cite{wan_quantum_2017}
\begin{equation}\label{Eq:Cost2}
C=\frac{1}{2}(\langle S_{z}^0\rangle_{des}^{ss}-\langle S_{z}^0\rangle_{act}^{ss})^2.
\end{equation}
Here, $\langle S_{z}^0\rangle_{des}^{ss}$ is the desired and $\langle S_{z}^0\rangle_{act}^{ss}$ is the actual steady magnetization values of the PQS for the spin observable $S_{z}$.
As expressed in Appendix C, the actual steady state magnetization of the PQS is a function of the coupling rates $g_i$. 
In this scheme, in analogy with Eq.~\ref{Eq:Weight}, the coupling rates are iterated as 
\begin{equation}\label{Eq:deltag}
g_{k+1}=g_k+\delta_{g_k}
\end{equation}
where 
\begin{equation}\label{Eq:DelC}
\delta g_k=-\eta\frac{\partial C}{\partial g_k}.
\end{equation}
Here, $\eta$ is the learning rate determining the variation speed of the relevant parameter in the direction of the greatest descent. 
(See Appendix C for the derivation of the cost function and its derivative for $g_i$ depending on the value of the $J$ parameter.)  

Figs.~\ref{fig:Epoch_Nd_1_tan} and \ref{fig:Epoch_Nd_1_2} illustrate the cost minimization against the iteration steps where the PQS is connected to two distinct information reservoirs with arbitrary couplings. 
In both Figures, we observe smooth convergence for cost minimization due to the continuous dynamics of the dissipation-assisted model. As obvious in Fig.~\ref{fig:Epoch_Nd_1_tan}, a learning rate of two orders of magnitude smaller than the coupling rate can reasonably minimize the cost function. Moreover, we observe that by keeping the learning rate and all the remaining parameters fixed, the speed of minimization increases with higher $J$, as shown in Fig.~\ref{fig:Epoch_Nd_1_2}.

\begin{figure}[!b]
\centering
\includegraphics[width=3.2in]{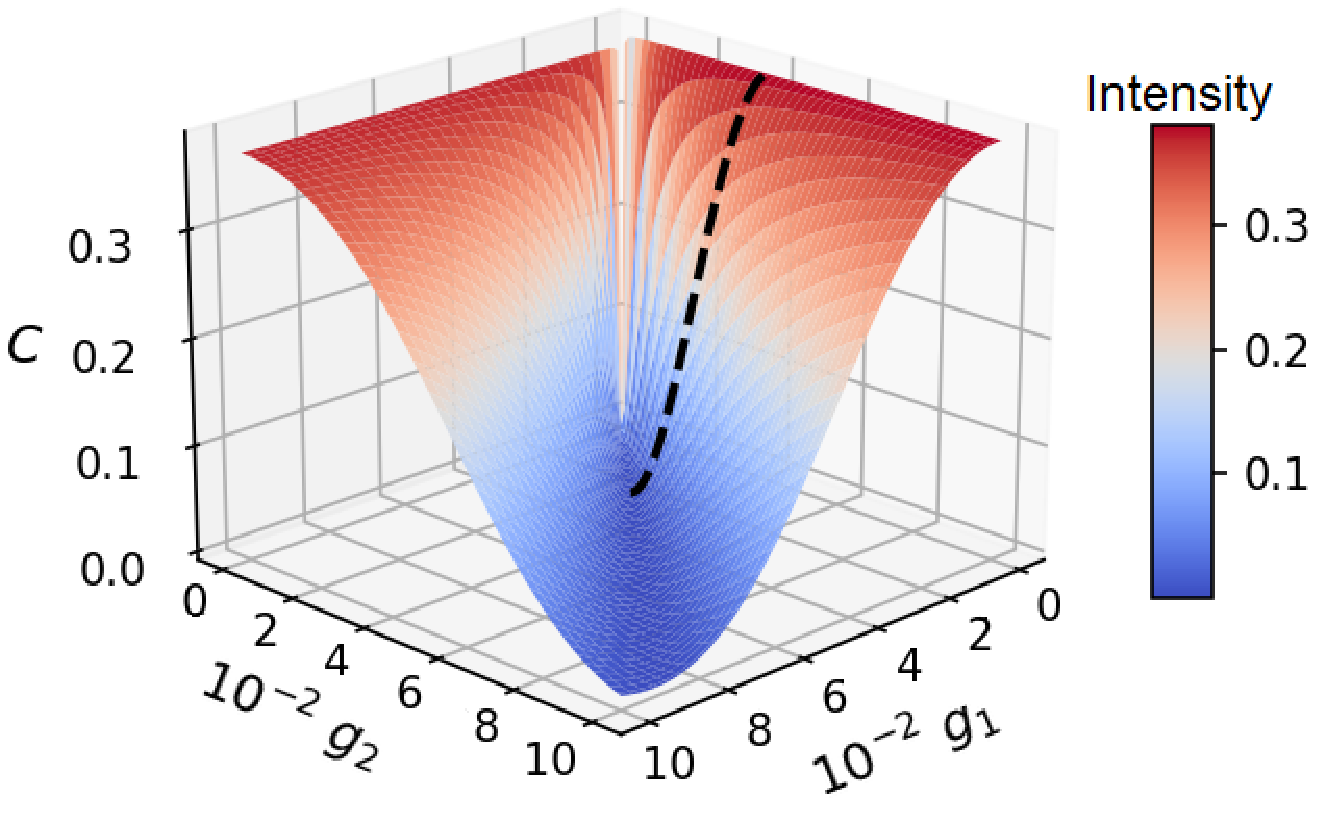}
\caption{ Cost function minimization with surface depiction against the variation of $g_1$ and $g_2$. The spin angular momentum value is $J=1$. The learning rate $\eta=2.4\times10^{-5}$, $\langle S_z^1\rangle=0.96$, $\langle S_z^2\rangle=-0.64$, 
the target magnetization is $\langle S_z^0\rangle_{des}^{ss}=0.12$ and the initial coupling rates before optimization is $g_1=0.001$, $g_2=0.04$.}
\label{fig:Cost_g_nd_2}
\end{figure}

The cost function minimization for a two-dimensional parameter space is depicted in Fig.~\ref{fig:Cost_g_nd_2}. The cost function has a smooth landscape without any local minima or plateaus. The optimization is achieved by updates in the coupling rate, as indicated by the dashed line. The results demonstrate that the model can successfully optimize the cost function without overshooting.

\section{Physical model}

The non-linear quantum neuron, which is represented by a d-dim. Hilbert space with total angular momentum $J>1/2$, has been introduced above as a theoretical model. Now, we aim to focus on its potential physical applications. As previously mentioned, the central idea of the model is based on a dynamic process that results from the repeated interaction of 2-level quantum systems (qubits) with a d-level PQS (qudit) system. In summary, we can describe this model as a quantum information processing that involves interactions between qubits and a qudit. The use of qudits in quantum information processing offers a larger Hilbert space due to its multi-level structure, allowing for richer information storage and processing with enhanced parallel processing capabilities~\cite{lu_quantum_2020, li_geometry_2013}. Despite receiving less attention compared to qubits, there are numerous natural and artificial physical applications related to qudits.

The physical systems with intrinsic angular momentum higher than $J=1/2$ are ubiquitous in nuclear spin systems or in subatomic particles~\cite{levitt_spin_2008}. However, they can also be obtained through effective Hamiltonians in set-ups such as trapped ions~\cite{ringbauer_universal_2022}, magnetic molecules~\cite{jenkins_coherent_2017}, or solid state systems~\cite{kobayashi_engineering_2021}. The emulation of a quantum spin system with a maximum angular momentum of $J=3/2$ has been achieved using superconducting quantum circuits~\cite{neeley_emulation_2009}.

Previous research has shown promising results in implementing qubit-qudit interactions efficiently through hybrid quantum systems, such as solid-state~\cite{amsuss_cavity_2011} or spin ensembles~\cite{schuster_high-cooperativity_2010,kubo_hybrid_2011} coupled to superconducting circuits. Most recently, the achievement of a [Yb(trensal)] molecular qudit with a nuclear spin of $I=5/2$ coupled to an effective electronic spin of $S=1/2$ through circuit quantum electrodynamics (c-QED) has been reported~\cite{rollano_high_2022}. The experimental architectures are constructed using lumped-element resonators, whose properties can be adjusted precisely to match the diverse nuclear and electronic spin transitions in the molecular system~\cite{aja_analysis_2021}. The studies mentioned above demonstrate that the model we presented is experimentally applicable.

\section{Conclusions}

In this study, we present a quantum model that has the potential to be the basic processing unit of a feedforward artificial neural network. The model operates within the framework of an open quantum system and is shown to be a differentiable, non-linear activation quantum model that can perform binary classification based on both numerical and analytical results. Our quantum neural model exhibits a response behavior similar to the well-known hyperbolic tangent function in the literature of classical artificial neural networks. Based on the results obtained from our model, we show that the non-linear behaviour is strengthened by the total angular momentum value $J$ of the system, which we refer to as PQS.

We obtained analytical results for the steady state using a master equation based on the collision model used for the description of the open quantum system, which allows for the determination of the input quantum information using the Bloch sphere parameter. Using the obtained results, we developed a cost function to test the applicability of the proposed model to learning processes. The cost function is shown to be continuously smooth and dependent on the value of $J$ in order to minimize it.

Furthermore, the current state-of-the-art already allows for the experimental implementation of small-scale models of the proposed quantum neural network. We believe that the model we propose is suitable for hybrid quantum computing processes and can find application in neuromorphic hardware-based quantum computers that operate in the NISQ regime.

\appendices
\section{The Unitary Propagator}\label{AppA}
The creation of the unitary propagator is the focus of our collision model-based analytical calculations.
First, we denote $U$ as the relevant Hamiltonian in the interaction picture.  
The unitary time-evolution operator is written as $\mathcal{U}(\tau)=\text{exp}[-\text{i}\mathcal{H}_{\text{int}}\tau]=\text{exp}[-\text{i}\tau U]$. Here, 
\begin{align}\label{Eq:U_Sum}
U=\sum_{i=1}^N g_i(S_0^{+}\sigma_i^{-}+\text{H.c.}) 
\end{align}
with $\hbar=1$. Then the propagator reads 
\begin{equation}\label{Eq:U_Approx}
\mathcal{U}(\tau)\simeq {\bf{1}} -\text{i}\tau U-\frac{\tau^2 U^2}{2}
\end{equation}
by evaluating the approximation up to second order in $\tau$ where $\text{i}=\sqrt{-1}$. 
$S_0^{\mp} = S_x \mp \text{i}S_y$ are the spin ladder operators acting on the PQS. The expression for the propagator can be written in matrix form for any value of the spin angular momentum $J$ of the PQS using Eq.~\ref{Eq:U_Approx}. Here, we only present the expression for $J=1$ for simplicity, with the analytical calculations for higher values of $J$ ($J\geq1$) given in the next appendix sections. The spin operators for $J=1$ read as
\begin{subequations}
\begin{align}
& S_x=\frac{1}{\sqrt{2}}
\begin{bmatrix}\label{Eq:S_x}
0  & 1 &0 \\
1 & 0 & 1\\
0& 1 & 0\\
\end{bmatrix}\\
& S_y=\frac{1}{\sqrt{2}}
\begin{bmatrix}\label{Eq:S_y}
0  & -\text{i}\ &0 \\
\text{i}\ & 0 & -\text{i}\\\
0& \text{i}\ & 0 \\
\end{bmatrix}\\
& S_z=
\begin{bmatrix}\label{Eq:S_z}
1 & 0 &0 \\
0 & 0 & 0\\
0& 0 & -1
\end{bmatrix}
\end{align}
\end{subequations}
and the matrices for $U$ and $U^2$ are as follows:
\begin{subequations}
\begin{align}
& U=
\begin{bmatrix}\label{Eq:U}
 & \mathcal{S}^{-}_{g_i} & \\
\mathcal{S}^{+}_{g_i} & & \mathcal{S}^{-}_{g_i}\\
& \mathcal{S}^{+}_{g_i} & \\
\end{bmatrix}\\
& U^2=
\begin{bmatrix}\label{Eq:U_Sq}
 \mathcal{S}^{-}_{g_i}\mathcal{S}^{+}_{g_i} & & \\
 &\mathcal{S}^{-}_{g_i}\mathcal{S}^{+}_{g_i}+\mathcal{S}^{+}_{g_i}\mathcal{S}^{-}_{g_i} & \\
 & & \mathcal{S}^{+}_{g_i}\mathcal{S}^{-}_{g_i} 
\end{bmatrix}
\end{align}
\end{subequations}
where $\mathcal{S}^{\pm}_{g_i}=\sum_{i=1}^N g_i\sigma_i^{\pm}$ represent the collective operators acting on the reservoir information units weighted by $g_i$. 
Therefore one obtains the corresponding propagator as
\begin{equation}
\mathcal{U}(\tau)=
\begin{bmatrix}
\bf{1}-\frac{\tau^2}{2}\mathcal{S}^{-}_{g_i}\mathcal{S}^{+}_{g_i} & -\text{i}\tau \mathcal{S}^{-}_{g_i} & \\
-\text{i}\tau\mathcal{S}^{+}_{g_i} & \bf{1}-\frac{\tau^2}{2}\mathcal{S}^{+}_{g_i}\mathcal{S}^{-}_{g_i}&-\text{i}\tau\mathcal{S}^{-}_{g_i}\\
 & -\text{i}\tau\mathcal{S}^{+}_{g_i}& \bf{1}-\frac{\tau^2}{2}\mathcal{S}^{+}_{g_i}\mathcal{S}^{-}_{g_i}
\end{bmatrix}.
\end{equation}

\section{The Master Equation}\label{AppB}

As stated in the text, our analytical results are obtained through the derivation of a master equation based on repeated interactions. According to the method we follow, repeated interactions can be represented as a random process described by the Poisson distribution. Therefore, the density matrix representing the entire system within a time interval $\delta t$ can be expressed as 
\begin{align} \label{Eq:Delta_t}
\varrho(t+\delta t)=r\delta t \mathcal{U}(\tau)\varrho(t)\mathcal{U}^{\dagger}(\tau)+(1-r\delta t)\varrho(t)
\end{align}
Here, $r\delta t$ represents the likelihood of an interaction occurring at a rate of $r$, while $1-r\delta t$ represents the likelihood of a non-interaction state. As $\delta t\rightarrow 0$, the following master equation 
\begin{align}\label{Eq:MicroMaser}
\dot{\varrho}_0(t)=r\text{Tr}_{\mathcal{R}_i}[\mathcal{U}(\tau)\varrho(t)\mathcal{U}^{\dagger}(\tau)-\varrho(t)]
\end{align}
for the reduced dynamics of PQS is derived from $\dot{\varrho}_0(t)=(\varrho_0(t+\delta t)-\varrho_0(t))/\delta t$.
Rewriting Eq.(\ref{Eq:U_Approx}) as $\mathcal{U}(\tau)\simeq {\bf{1}}-U_1(\tau)-U_2(\tau)$ with $U_1(\tau)=\text{i}\tau U$ and $U_2(\tau)=\tau^2 U^2/2$, and inserting into Eq.(\ref{Eq:MicroMaser}), one obtains 
\begin{align}
\dot{\varrho}_0(t)=& r\text{Tr}_{\mathcal{R}_i}[U_1(\tau)\varrho(t)U_1^{\dagger}(\tau)-U_1(\tau)\varrho(t)\nonumber\\
&-U_2(\tau)\varrho(t)-\varrho(t) U_1^{\dagger}(\tau)-\varrho(t) U_2^{\dagger}(\tau)]. 
\end{align}
Here, we have disregarded terms that are higher than second order in $\tau$, such as $U_2(\tau)\varrho(t) U_2^{\dagger}(\tau)\propto \tau^4$ or $U_1(\tau)\varrho(t) U_2^{\dagger}(\tau)\propto \tau^3$. The explicit form of the master equation for PQS is obtained by tracing out the degrees of freedom in the information environment and utilizing the linear and cyclic properties of the trace operation. 

Following the recipe above, the obtained master equation for the proposed model reads 
\begin{multline}
\dot{\varrho}_0=-\text{i}r\tau\left[\sum_i^N g_i\left(\langle\sigma_i^{-}\rangle S_0^{+}+\langle\sigma_i^{+}\rangle S_0^{-}\right),\varrho\right]\\
+\frac{r\tau^2}{2}\sum_{i=1}^N g_i^2\left(\langle \sigma_i^{+}\sigma_i^{-}\rangle\mathcal{L}[S_0^{+}]+\langle \sigma_i^{-}\sigma_i^{+}\rangle\mathcal{L}[S_0^{-}]\right)\\
+2r\tau^2\sum_{i<j}^{N'} g_i g_j \left(\langle \sigma_i^{+}\rangle\langle\sigma_j^{+}\rangle \mathcal{L}_s[S_0^{-}]+\langle \sigma_i^{-}\rangle\langle\sigma_j^{-}\rangle \mathcal{L}_s[S_0^{+}]\right).
\end{multline}

The above expression provides the mathematical proof of the transfer of information from the reservoirs to PQS through the calculation of average values obtained with Pauli operators. The relationship between these average values and the density matrix elements of the information reservoirs, which represent the input quantum information, is related to the relevant parametrizations as 
\begin{align}\label{Eq:RhoR}
\mathcal{\rho}_{\mathcal{R}_{i}}&=
\begin{bmatrix}
\frac{1+\cos\theta_{i}}{2} & \frac{e^{-i\phi_{i}}}{2}\sin \theta_{i} \\
\frac{e^{i\phi_{i}}}{2}\sin \theta_{i} & \frac{1-\cos\theta_{i}}{2}
\end{bmatrix}:=
\begin{bmatrix}
\langle\sigma_{i}^+\sigma_{i}^-\rangle & \langle\sigma_{i}^-\rangle \\
\langle\sigma_{i}^+\rangle & \langle\sigma_{i}^-\sigma_{i}^+\rangle
\end{bmatrix}
\end{align}
where $\mathcal{\rho}_{\mathcal{R}_{i}}$ is the $i$th information reservoir density matrix. 

The density matrix representation of the initial PQS state can be stated as
\begin{equation}\label{Eq:RH0} 
\varrho_0=\frac{1}{3}\begin{bmatrix} P_{11} & C_{12} & C_{13} \\
C_{21} &  P_{22} & C_{23} \\
C_{31} & C_{32} & P_{33} \end{bmatrix}\equiv \ket{PQS}_{0~0}\bra{PQS}
\end{equation}
where $\ket{PQS}_0=\frac{1}{\sqrt{3}}(c_1\ket{e_1}+c_2\ket{e_2}+c_3\ket{g})$ with $\bra{e_1}=(1\quad 0\quad 0)$,
$\bra{e_2}=(0\quad 1\quad 0)$, and $\bra{e_3}=(0\quad 0\quad 1)$ being the dual form of the eigenbasis  of $S_z$ for $J=1$. Note that  $P_{nn} =c_n c_n^*$ and $C_{nm} =c_n c_m^*$. Following these notations, the corresponding Bloch equations read
\begin{align}\label{Eq:Bloch_nd_2}
\langle{\dot{S_{x}}}\rangle=&\frac{1}{3\sqrt{2}}\left( 2\gamma^-_3 C_{21}-\gamma^+_2 C_{21}+2\gamma^+_3 C_{12}-\gamma^-_2 C_{23}+2\gamma^-_3 C_{32}\right) \nonumber\\
&+\frac{1}{3\sqrt{2}} \left(-\gamma^-_2 C_{32}+2\gamma^+_3 C_{23}\right)+\frac{{\langle S_{z}\rangle}_0}{\sqrt{2}}\left( \gamma^-_1-\gamma^+_1 \right)\\
\langle{\dot{S_{y}}}\rangle=&\frac{\text{i}}{3\sqrt{2}}\left( 2\gamma^-_3 C_{21}+\gamma^+_2 C_{21}-2\gamma^+_3 C_{12}-\gamma^-_2 C_{23}+2\gamma^-_3 C_{32}\right) \nonumber\\
&+\frac{i}{3\sqrt{2}} \left(\gamma^-_2 C_{32}-2\gamma^+_3 C_{23}\right)+i\frac{{\langle S_{z}\rangle}_0}{\sqrt{2}}\left( \gamma^-_1+\gamma^+_1 \right)\\
\langle\dot{S_{z}}\rangle=&\gamma^+_1  C_{12}-\gamma^-_1  P_{22}+2\gamma^+_2 \left( P_{22}+P_{33}\right)-2\gamma^-_2 \left( P_{11}+P_{22}\right)\nonumber\\
&-\gamma^-_1 C_{32}+i\gamma^+_1 C_{23}).
\end{align}

As previously stated, the presented model gives the binary classification decision in a steady state. Therefore, the steady state solutions of the master equation and the Bloch equations should be obtained. The steady state of the PQS density matrix is expressed in Eq.~\ref{Eq:Steadyy} as the solution of $\dot{\varrho}_0=0$. Solutions of the Bloch equations in the steady state found to be as
\begin{align}\label{Eq:Bloch_ss_nd_2}
&\langle S_{z}\rangle^{ss}=3{\langle S_{z}\rangle}_0\nonumber\\
&=\frac{(\sum_{i=1}^N g^2_i p_{e_i})^2-(\sum_{i=1}^N g^2_i p_{g_i})^2}{(\sum_{i=1}^N g^2_i p_{e_i})^2+(\sum_{i=1}^N g^2_i p_{g_i})^2+\sum_{i,j=1}^N g^2_i g^2_j p_{e_i}p_{g_j}}\nonumber\\
&\langle S_{z}\rangle^{ss}=\frac{g_{1}^{4}\langle \sigma_z^1\rangle+g_{2}^{4}\langle \sigma_z^2\rangle+2g_{1}^{2}g_{2}^{2}(p_{e_1}p_{e_2}-p_{g_1}p_{g_2})}{g_{1}^{4}(1-p_{e_1}p_{g_1})+g_{2}^{4}(1-p_{e_2}p_{g_2})+g_{1}^{2}g_{2}^{2}(1+p_{g_1}p_{g_2})}\\
&\langle S_{y}\rangle^{ss}=\text{i}\frac{3\langle S_{z}\rangle_{0}}{\sqrt{2}}\left( \gamma^-_1+\gamma^+_1 \right)\\
&\langle S_{x}\rangle^{ss}= \frac{3\langle S_{z}\rangle_{0}}{\sqrt{2}}\left( \gamma^-_1-\gamma^+_1 \right)
\end{align}
where ${\langle S_{z}\rangle}_0=\frac{1}{3}(P_{11}-P_{33})$, and $p_{e_i}$ and $p_{g_i}$ are, the diagonal elements of the reservoir units' density matrix.  

The expressions for the steady magnetization and the Bloch solutions of PQS can be obtained using the same methods by considering higher values of $J$ in Eq.~\ref{Eq:U_Approx}. Unfortunately, the expressions obtained with higher spin values are too complicated to be generalized for a general $J$ value, so we will only share the graph we obtained for the cost function using the analytical results we obtained up to $J=5/2$. However, as stated in the main text, higher values of $J$ for PQS result in a stronger non-linear response in the steady state and provide a faster decrease in the cost function, which will be further discussed below.

\begin{figure}[!t]
\centering
\includegraphics[width=3.4in]{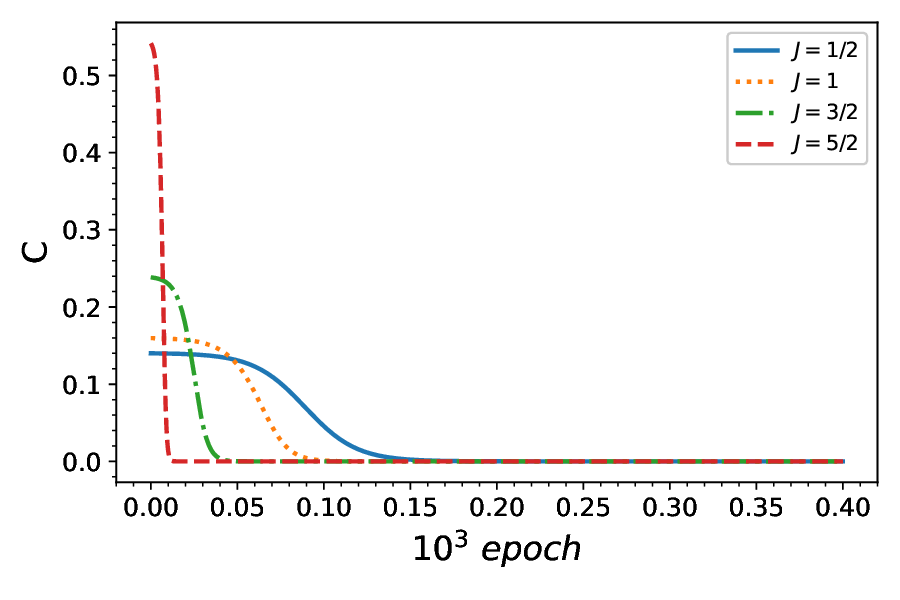}
\caption{ Cost function minimization against iteration steps depending on different values of $J$. The spin angular quantum numbers are $J=1/2$ for the blue solid line, $J=1$ for the orange dotted line, $J=3/2$ for the green dash-dotted line, and $J=5/2$ for the red dashed line. For the both of them the initial values of the reservoirs are  $\langle \sigma_z^1\rangle=0.95$, $\langle \sigma_z^2\rangle=-0.11$, $g_1=0.001$, $g_2=0.04$ and $\langle S_z^0\rangle_{des}^{ss}=0.42$, respectively.}
\label{fig:Cost_J}
\end{figure}

\section{Derivation of the cost function}\label{AppC}

In this section, we analyze the derivation of the cost function using Eqs.~(\ref{Eq:Cost2}),~(\ref{Eq:deltag}) and (\ref{Eq:DelC}). To start, we extend Eq.~(\ref{Eq:DelC}) by taking the partial derivative with respect to $g_i$ as 
\begin{equation}\label{Eq:deltac}
\frac{\partial C}{\partial {g}_i}=(\langle S_z^0\rangle_{des}^{ss}-\langle S_z^0\rangle_{act}^{ss})(-\frac{\partial \langle S_z^0\rangle_{act}^{ss}}{\partial {g}_i}).
\end{equation}
It's important to note that in the current example, the total angular momentum of the probe system is $J=1$ and PQS is connected to two information reservoirs. As a result, we use the steady magnetization, represented by $A$, as the actual value of the cost function as
 \begin{align}\label{Eq:Actual}
A&=\langle S_z^0\rangle_{act}^{ss}\nonumber\\
&=\frac{g_{1}^{4}\langle \sigma_z^1\rangle+g_{2}^{4}\langle \sigma_z^2\rangle+2g_{1}^{2}g_{2}^{2}(p_{e_1}p_{e_2}-p_{g_1}p_{g_2})}{g_{1}^{4}(1-p_{e_1}p_{g_1})+g_{2}^{4}(1-p_{e_2}p_{g_2})+g_{1}^{2}g_{2}^{2}(1+p_{g_1}p_{g_2})}.
\end{align} 

Here, $\langle \sigma_z^i\rangle=p_{e_i}-p_{g_i}$. For simplicity, we denote the numerator in eq.~(\ref{Eq:Actual}) as $X(g_1,g_2)$ and the denominator as $Y(g_1,g_2)$. According to the recipe to derive the cost function, the partial derivatives with respect to $g_1$ and $g_2$ separately obtained as
\begin{align}\label{Eq:pdg1_g2}
&\frac{\partial A}{\partial {g}_1}=\frac{\frac{\partial X(g_1,g_2)}{\partial {g}_1}Y(g_1,g_2)-\frac{\partial Y(g_1,g_2)}{\partial {g}_1}X(g_1,g_2)}{{(Y(g_1,g_2))}^{2}}\nonumber\\
&\frac{\partial A}{\partial {g}_2}=\frac{\frac{\partial X(g_1,g_2)}{\partial {g}_2}Y(g_1,g_2)-\frac{\partial Y(g_1,g_2)}{\partial {g}_2}X(g_1,g_2)}{{(Y(g_1,g_2))}^{2}}
\end{align}
In our example, the desired magnetization $\langle S_z^0\rangle_{des}^{ss}=0.4$ is an arbitrary value in the cost function. By substituting Eqs.~(\ref{Eq:Actual}) and (\ref{Eq:pdg1_g2}) into Eq.~(\ref{Eq:deltac}), and Eq.~\ref{Eq:DelC} into Eq.~\ref{Eq:deltag} we obtain the explicit form of the cost function and the coupling strengths update expressions
as
\begin{align}\label{Eq:training_1}
&(g_1)_{k+1}=(g_1)_{k}+\delta(g_1)_{k}\nonumber\\
&(g_2)_{k+1}=(g_2)_{k}+\delta(g_2)_{k}.
\end{align}

In the main text, we compared the cost function in Fig.\ref{fig_epoch} for $J=1/2$ and $J=1$, and in Fig.\ref{fig:Cost_J}, we have compared it for higher values of $J$. Since analytical expressions become cumbersome for higher values of $J$, we only provide the graphical results as shown Fig.~\ref{fig:Cost_J}. The obtained results show that for the same learning rate values, changing only the $J$ value rapidly affects the convergence in the iteration.

\section*{Acknowledgments}
The authors acknowledge support from the Scientific and Technological Research Council of Turkey (TÜBİTAK-Grant No. 120F353). The authors also wish to extend special thanks to the  Cognitive Systems Lab in the Department of Electrical Engineering providing the atmosphere for motivational and stimulating discussions.


\newpage

\begin{IEEEbiography}[{\includegraphics[width=1in,height=1.25in,clip,keepaspectratio]{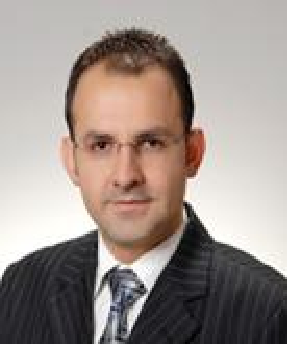}}]{Ufuk Korkmaz}
received the BSc, MSc and PhD degrees from the Ondokuz Mayıs University (OMU), Physics Department, in 2006, 2010 and 2014 respectively. He worked as a Postdoctoral Researcher at Istanbul Technical University (ITU) in 2018-2019. He starting researches as Post-Doc in Istanbul Technical University (ITU) in 2021. His research interests are IR and UV spectroscopy, X-ray single crystal diffraction, Understanding the nature of H bonds in supramolecular structure, Quantum Mechanics and Quantum information theory. 
\end{IEEEbiography}
\vspace{11pt}
\begin{IEEEbiography}[{\includegraphics[width=1in,height=1.25in,clip,keepaspectratio]{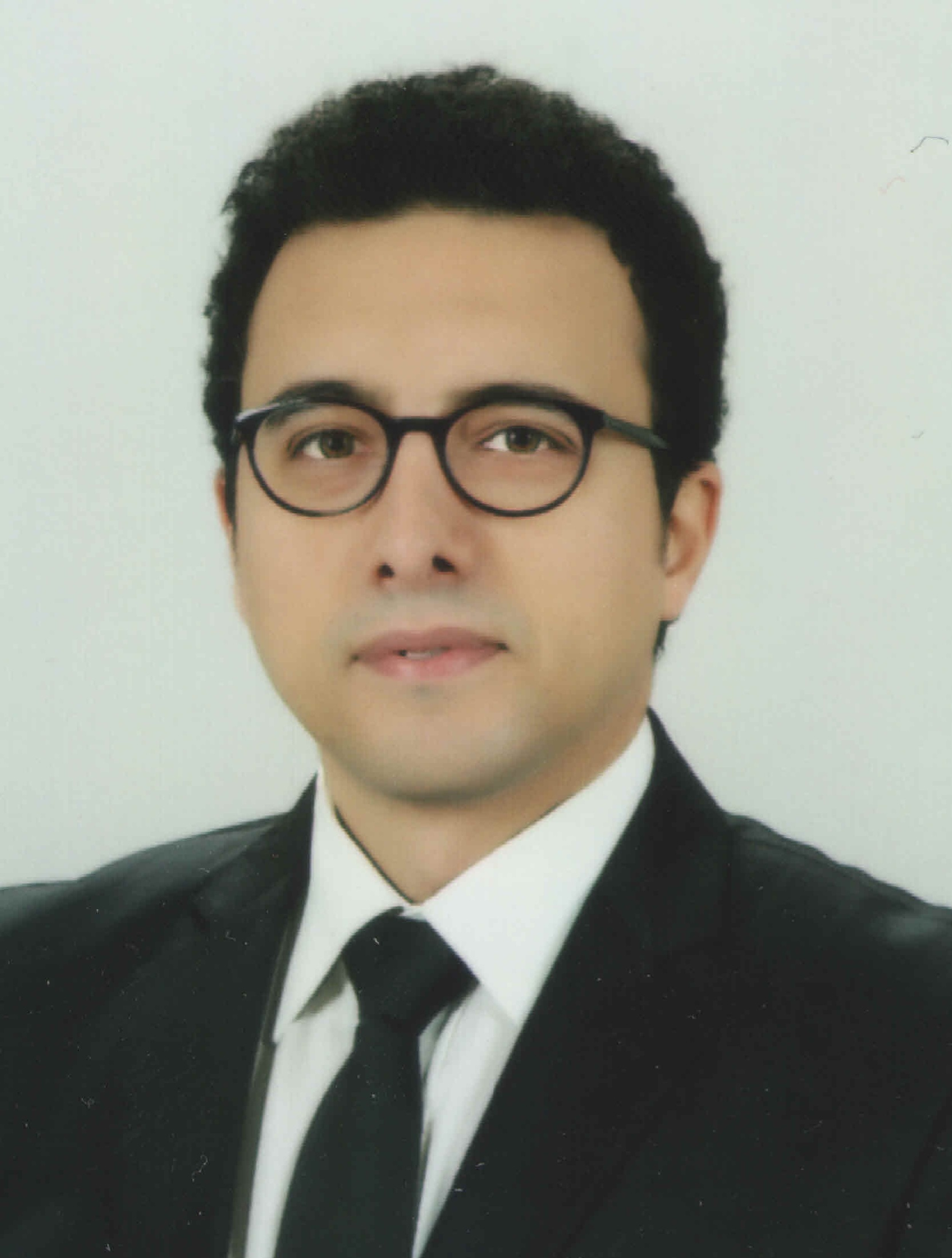}}]{Deniz Türkpençe}
received M.S., and Ph.D. degrees in Atomic Physics from from Ondokuz Mayıs University in 2007 and 2013. He worked as a visiting student researcher at Dortmund Technical University for 1 year with a YÖK scholarship to conduct research and examination abroad related to his doctoral thesis. He worked as a Research Assistant at Koç University in April 2014 and took part in an international project. He completed her studies at Koç University in 2016. He worked as a postdoctoral researcher in the Cognitive Systems Laboratory of the Electrical Engineering Department of Istanbul Technical University between 2017-2018. He started to work as a  Lecturer in the Department of Electrical Engineering at Istanbul Technical University in March 2018. In November 2019, he was awarded the title of Associate Professor by the Turkey Inter-university Board. Currently he is a faculty member in Electric Electronic Faculty of Istanbul Technical University. 
\end{IEEEbiography}
\vfill
\end{document}